# Synthesis and properties of Co-doped titanate nanotubes and their optical sensitization with methylene blue

V.C. Ferreira[1], M.R. Nunes[2], A.J. Silvestre[3] and O.C. Monteiro[1*]

[1] University of Lisbon, Faculty of Sciences, Department of Chemistry and Biochemistry and CQB, Campo Grande, 1749-016 Lisboa, Portugal

[2] University of Lisbon, Faculty of Sciences, Department of Chemistry and Biochemistry and CCMM, Campo Grande, 1749-016 Lisboa, Portugal

[3] Instituto Superior de Engenharia de Lisboa, Department of Physics and ICEMS, R. Conselheiro Emídio Navarro 1, 1959-007 Lisboa, Portugal

**Abstract**

Here we report on a novel chemical route to synthesize homogenous cobalt doped titanate nanotubes (CoTNT), using an amorphous Co-doped precursor. The influence of the synthesis temperature, autoclave dwell time and metal doping on the structural and microstructural as well as on the optical properties of the synthesized titanate nanotubes is studied and discussed. The optical band gaps of the CoTNT samples are red shifted in comparison with the values determined for the undoped samples, such red shifts bringing the absorption edge of the CoTNT samples into the visible region. CoTNT materials also demonstrate particular high adsorption ability for methylene blue, the amount of the adsorbed dye being higher than the one predictable for a monolayer formation. This suggests the possibility of intercalation of the dye molecule between the $TiO_6$ layers of the TNT structure. It is also shown that the methylene blue sensitized Co-doped nanostructures are highly stable under UV radiation and present a strong and broad absorption in the visible region.

---

[1] Corresponding author: Phone:+351 217500865; Fax:+351 217500088; E-mail: ocmonteiro@fc.ul.pt

# 1. Introduction

Stimulated by the discovery of carbon nanotubes, one-dimensional nanostructured materials have become an intense research topic in nanotechnology owing to their unusual properties and a wide variety of potential applications. Among them, titanate nanotubes (TNTs) have been subject of vast interest because of their open mesoporous morphology with high specific surface area, good ion-exchange properties and electrical conductivity, and cheap fabrication [1,2]. For a review of the wide potential applications of the TNT-*based* materials see *e.g.* the recent book of Bavykin and Walsh [3].

The walls of the titanate nanotubes have a characteristic multilayered structure consisting of edge- and corner-sharing $TiO_6$ octahedra building up zigzag structures with the sodium ions located between the $TiO_6$ layers. The detailed crystal structure of TNT is still controversial, although it has been assumed that the TNTs are formed via scrolling of nanosheets exfoliated from lamellar titanates whose structures have been assigned to wide variety of possible compositions such as $Na_2Ti_3O_7$, $H_2Ti_3O_7$, $Na_xH_{2-x}Ti_3O_7$, $Na_xH_{2-x}Ti_3O_7 \cdot nH_2O$, $Na_2Ti_2O_4(OH)_2$, $H_2Ti_2O_4(OH)_2$, $Na_2Ti_2O_5 \cdot H_2O$, $H_2Ti_2O_5 \cdot H_2O$ or even the *lepidocrocite*-type titanates with compositions $Na_xTi_{2-x/4}\square_{x/4}O_4$ or $H_xTi_{2-x/4}\square_{x/4}O_4$, where $\square$ stands for vacancy [4-8].

The TNTs' ion-exchange ability makes them potential materials for effective photocatalysts and solar energy cells applications [9-11]. However, they present a critical drawback: their high charge recombination rate and wide band gap (*ca.* 3.3 eV) limit the electron and hole photo-generation under visible irradiation. Therefore, the synthesis of TNT-based materials with either a broader range of light absorption and/or a lower charge recombination rate would be an important achievement toward the development of successful photoactive materials. Several works have been published on this issue including metal doping [12,13], co-sensitized with semiconductor nanoparticles and porphyrin zinc co-sensitization [14] or phthalocyanin-sensitization [15].

This work reports on a swift chemical route to synthesize homogenous cobalt doped titanate nanotubes (CoTNTs), extending the hydrothermal process used to prepare undoped TNTs previously reported by our group [9]. It is shown that Co doping stabilizes the morphology of the TNTs in a synthesis temperature range of 160 ºC to 200 ºC, and brings TNTs' absorption edge to the visible region. The adsorption ability of the synthesized materials for methylene blue (MB) is studied and it is shown that not only the MB-sensitized CoTNT materials are highly stable under UV irradiation but also they present a strong and broad absorption in the visible region of the electromagnetic spectrum.

## 2. Materials and Methods

All reagents were of analytical grade (Aldrich and Fluka) and were used as received. The solutions were prepared with Millipore Milli-Q ultra-pure water.

### 2.1. Materials

#### 2.1.1. TNTs precursor synthesis

The TNT precursor was prepared using a procedure previously reported by our group [16]. The used titanium source was a titanium trichloride solution (10 wt.% in 20-30 wt.% HCl) diluted in a ratio of 1:2 in standard HCl solution (37 %). To this solution an ammonia solution 4 M was added dropwise under vigorous stirring, until a complete precipitation of a white solid was observed. The resulting suspension was kept overnight at room temperature and then filtered and washed with deionised water. The as-prepared precursor shows an amorphous structure and its characterization has been reported recently [9].

An identical synthesis route was used to synthesise a Co-doped TNT precursor, by adding the required molar amount (5 %) of cobalt to the titanium trichloride solution using the following procedure: 0.170 g of metallic cobalt powder (Johnson Matthey) was carefully dissolved in *ca*. 1 mL of concentrated HNO$_3$. The deep purple solution was heated to evaporate the solvent and *ca*. 2 mL of HCl (37 %) solution was added to evaporate residual nitrates (yellow steam is formed). After no yellowish steam was observed, the solution was diluted with 2 M HCl to 100 mL final volume. This solution was added to the 50 mL of the titanium trichloride solution. After precipitation with the 4 M ammonia solution, the CoTNT precursor was obtained as a greenish grey solid.

*2.1.2. TNT and CoTNT synthesis*

The undoped and Co-doped samples, respectively TNT and CoTNT, were prepared into an autoclave system using ~6 g of precursor in *ca.* 60 mL of NaOH 10 M aqueous solution. The samples were prepared at 160 ºC and 200 ºC using autoclave dwell times of 12 h and 24 h, respectively. After cooling, the suspensions were filtered and the solids white (TNT) and grey (CoTNT), were washed systematically. Each solid was dispersed in ~150 mL of water and magnetically stirred for one hour. Then, the suspension was filtered and the *pH* of the filtrate measured for reference. This procedure was repeated until filtrate solutions have reached *pH* = 7. The solids were afterwards dried and stored. For the sake of clarity, the undoped samples were labelled as "TNT'temperature'time" and the Co-doped samples as "CoTNT'temperature'time"; e.g. the TNT16024 and CoTNT16024 samples are the undoped and Co-doped samples prepared at 160 ºC during 24 h, respectively.

*2.1.3. TNT and CoTNT sensitization*

The sensitization/adsorption studies were carried out using Methylene Blue (MB) as a probe dye. Samples were prepared as follows: 10 mg of TNT or CoTNT were suspended in a dye aqueous solution (250 mg L$^{-1}$), and stirred for 2 h in dark conditions. After centrifugation, the MB concentration was estimated by measuring the absorbance of the solution at 665 nm (MB's chromophoric peak).

*2.1.4. Electrochemical measurements*

Electrochemical measurements were carried out with a computer-controlled CHI620A electrochemical workstation in a conventional three-electrode system using a platinum foil counter electrode, a saturated calomel reference electrode (SCE) and a glassy carbon working electrode (GCE, geometric area $A_{GCE}$ = 0.196 cm$^2$). The GCE were modified with the TNT or CoTNT samples, previously saturated with MB, by a drop casting method (30 μL of a 1 mg mL$^{-1}$ suspension). After modification, the electrodes were dried overnight and characterized electrochemically by cyclic voltammetry. A 0.1 M NaClO$_4$ solution was used as supporting electrolytic solution and the system was deoxygenated for *ca.* 20 minutes with a nitrogen flow before measurements. The cyclic voltammograms were obtained at a scan rate of 50 mV s$^{-1}$.

*2.3. Characterization*

X-ray powder diffraction was performed using a Philips X-ray diffractometer (PW 1730) with automatic data acquisition (APD Philips v3.6B), using Cu Kα radiation ($\lambda$ = 0.15406 nm) and working at 40 kV/30 mA. The diffraction patterns were collected in the range $2\theta$ = 7°-60° with a 0.02° step size and an acquisition time of 2.0 s/step. Optical characterization of the samples was carried out by UV-Vis diffuse reflectance using a Shimadzu UV-2600PC spectrophotometer. Diffuse reflectance spectra (DRS) were recorded in the wavelength range of 220-1400 nm using an ISR 2600plus integration sphere. Transmission electron microscopy (TEM) and high resolution transmission electron microscopy (HRTEM) were carried out using a JEOL 200CX microscope operating at 200 kV. Specific surface areas were estimated by the BET method, from nitrogen (Air Liquide, 99.999%) adsorption data at −196 °C, using a volumetric apparatus from Quantachrome mod. NOVA 2200e. The samples, weighing between 40 and 60 mg, were previously degassed for 2.5 h at 150 °C at a pressure lower than 0.133 Pa.

## 3. Results and discussion

### 3.1. Structural and morphological characterization

In a previous work [9], we have shown that homogeneous undoped titanate nanotubes with a high length/diameter aspect ratio can be prepared at 160 ºC and an autoclave dwell time of 24 h, via the hydrothermal synthesis approach described here. Figure 1a shows the XRD pattern of an undoped sample synthesised in those conditions (sample TNT16024). As can be seen, diffraction peaks at $2\theta$ around 10.3º, 24.5º, 28.6º and 48.6º can be clearly identified and are in agreement with numerous reported XRD patterns assigned to titanate layered structures [4-8,10]. The broad peak at 10.3º is related with the TNTs interlayer distance [8,10]. The peak at 24.5º can be associated with the presence of hydrogen trititanate (JCPDS-ICDD file nº 41-0192), which reveals that $Na^+/H^+$ exchange in the TNTs interlayer region has occurred to some extent due to the sample washing process. Considering these facts, the diffractogram of Figure 1a cannot thoroughly confirm the structure of the synthesized TNTs. However, similar XRD patterns have been associated either to $Na_xH_{2-x}Ti_3O_7$ [17,18] or $Na_xH_{2-x}Ti_3O_7 \cdot nH_2O$ [8] structures. The microstructure of this sample can be seen in the TEM image in Figure 1b. As shown, the TNT16024 sample consists of very thin and homogeneous nanotubular structures with a high length/diameter aspect ratio. The TNTs' mean diameter, $<d>$, was determined by TEM image analysis. A mean value $<d> = 8 \pm 2$ nm was estimated for the sample TNT16024 (see inset of Figure 1b). In order to test the robustness of our synthesis procedure to prepare Co-doped TNT samples, a Co-doped sample using the same temperature/time routine (160 ºC/24 h) was prepared. Figure 1c shows the diffractogram of the sample prepared in these experimental conditions (sample CoTNT16024). As can be seen, the diffractogram is analogous to the one obtained from the undoped sample. Furthermore, there is no sign of undesirable phases, *e.g.* cobalt clusters, cobalt oxides or Co-Ti oxide phases, which are known to exist in the bulk Co-Ti-O phase diagram [19]. Note that the presence of the Co in addition to the Ti, O and Na elements was confirmed in this sample by EDS, with a Co:Ti ratio close to the nominal stoichiometry of 5% (see supplementary data, Figure S1). Therefore, the XRD pattern seems to provide evidence for the homogeneous distribution of the doping element, within the detection limit of the technique. The TEM image of sample CoTNT16024 (Figure 1d) reveals a microstructure identical to the one observed for the correspondent undoped TNT16024 sample, with TNTs mean diameter of $<d> = 7.1 \pm 0.8$ nm (see inset of Figure 1d). These results seem to attest the robustness of our synthesis approach to prepare Co-doped TNTs, and reported for the first time here. Concerning the temperature sensitivity of the synthesis processes, it is well known that variations of only a few tens of degrees in the alkaline hydrothermal chemical routes may lead to strong variations in the morphology of the synthesised nanostructures, including length, diameter and size of agglomerates [3]. In order to assess the temperature sensitivity of the Co-doped TNT synthesis procedure, undoped and Co-doped TNTs samples were prepared at 200 ºC for 12 h and compared with the ones prepared at 160 ºC. Figure 2a shows the XRD pattern of the TNT20012 sample. Beyond the diffraction peak at $2\theta$ around 10º, which is related with the TNT interlayer distance, the pattern reveals a material with a higher degree of crystallinity with several peaks that could be indexed to both the $Na_2Ti_3O_7$ and the $H_2Ti_3O_7$ JCPDS-ICDD files nº 31-1329 and 41-0192, respectively. Note that the existence of both titanate phases supports the existence of some $Na^+ \rightarrow H^+$ replacement in the TNTs structure induced by the samples washing process, as mentioned above. The TNT20012 sample morphology is shown in Figure 2b. As can be seen, it presents a very heterogeneous nanorod-*like* structure, with a mixture of nano and micro scale diameters of mean value $<d> = 110 \pm 81$ nm (see inset of Figure 2b). Surprisingly, the structure and morphology of the Co-doped sample synthesised at 200 ºC for 12 h (CoTNT20012) is comparable to the ones prepared at 160 ºC (CoTNT16024 and TNT16024). Its diffractogram (Figure 2c) is similar to the XRD patterns of samples TNT16024 and CoTNT16024 and its morphology consists of homogeneous nanotubular structures with a high length/diameter aspect ratio (Figure 2d), and mean diameter <d> = 7 ± 2 nm (see inset of Figure 2d). Therefore, beyond attesting the robustness of the synthesis approach to prepare Co-doped TNTs as previously mentioned, these results also suggest that Co-doping stabilizes the titanate nanotubes' morphology in a wider synthesis temperature range, allowing a significant reduction of the autoclave dwell time and thus a cost saving in energy consumption. The mechanism by which the morphology stabilization occurs is currently unknown, although it may be related with the ionic radius size effect of the $Co^{2+}$ ($r$ = 0.745 Å) when replacing the $Ti^{4+}$ ($r$ = 0.605 Å) into the $TiO_6$ octahedra TNT based blocks, and consequent increase of oxygen vacancies due to the unbalanced electrical charge caused by the $Ti^{4+} \rightarrow Co^{2+}$ ion substitution. Nevertheless, more studies are needed to clarify such mechanism.

**3.2. UV-vis optical response**

The optical characterization of the samples was carried out by measuring their diffuse reflectance, $R$, at room temperature. $R$ can be related with the absorption Kubelka-Munk function, $F_{KM}$, by the relation $F_{KM}(R) = (1-R)^2/2R$, which is proportional to the absorption coefficient [20]. The obtained absorption spectra are shown in Figure 3. The optical band gap energies of the samples ($E_g$) were estimated by plotting the function $f_{KM} = (F_{KM}h\nu)^{1/2}$ vs. radiation energy (Tauc plot), where $h$ stands for the Planck constant and $\nu$ for the radiation frequency, and by extrapolating the linear portion of the curve to zero absorption (Figure 4). The estimated $E_g$ values for the TNT16024 and TNT20012 undoped samples were 2.96 ± 0.03 eV and 3.34 ± 0.03 eV, respectively, while for the CoTNT16024 and CoTNT20012 Co-doped samples the values 2.37 ± 0.05 eV and 2.36 ± 0.04 eV were respectively inferred. Note that the optical band gap energies of the Co-doped TNTs are clear red shifted in relation to the ones obtained for the undoped TNTs, with absorption edges in the near-visible region. Moreover, their $E_g$ values are similar and independent of the synthesis temperature and autoclave dwell time tested in this work. Consequently, the red shift of the optical band gap deduced for the CoTNT samples seems to result from the Co doping process. Moreover, this result seems to be consistent with the hypothesis that the dopant element is homogeneously distributed in substitutional sites of the $TiO_6$ octahedra TNT building structures, and thus giving rise to $Ti_{1-x}Co_xO_{6-\delta}$ doped structures with oxygen vacancies due to the lower valence states of cobalt ($Co^{2+}$) in comparison with titanium ($Ti^{4+}$). Indeed, similar red shifts have been reported for other transition metal doped-$TiO_2$ based materials, including doping with V, Cr, Fe, Co, Ni, and Cu [12,21,22]. In particular, red shifts of Co-doped $TiO_2$ are consistent with the introduction of electronic states in the titanium oxide band gap associated with the 3$d$ electrons of $Co^{2+}$ cations and oxygen defects [23]. Actually, previous band-structure calculations have shown that the valence band derives primarily from O 2$p$-levels, the conduction band from the Ti 3$d$-levels, and that the crystal-field split Co 3$d$-levels form localized bands within the original band gap of $TiO_2$ [24-26]. Consequently, the optical absorption edge transition for the Co-doped $TiO_2$ results from the $n$-type doping of the $TiO_2$ matrix [23]. The results here reported seem to support that comparable effects may occur when Co substitutes Ti in the $TiO_6$ octahedra TNT building blocks. The localized states induced by Co doping are expected to form tails of states that extend the bands into the band gap, producing an absorption tail known as Urbach tail [27]. The Urbach energy, $E_U$, associated to the width of the Urbach tail follows the exponential law $\alpha = \alpha_0 \exp(h\nu/E_U)$, where $\alpha$ is the optical absorption coefficient and $\alpha_0$ is a constant [27]. Since the absorption Kubelka-Munk function $F_{KM}$ is proportional to the sample's absorption, the Urbach energies of the different TNT synthesized samples were estimated from the slopes of $\ln(F_{KM})$ plotted as a function of photon energy, in an energy range just below the gap (Figure 5a). The calculated $E_U$ values are given in Table 1 and the $E_g$ *versus* $E_U$ values are plotted in Figure 5b. As can be seen from Figure 5b, as the Urbach energy increases, the optical band gap energy decreases, increasing the red shift from possible band-to-tail and tail-to-tail transitions. The higher $E_U$ values are associated with the Co-doped samples, indicating further introduction of tails in the band gap of CoTNTs samples when compared with the undoped ones.

**3.3. Dye sensitization**

In addition to other potential applications, titanate elongated nanostructures are seen as promising materials for photovoltaic and solar applications. Therefore, the search of new processes for their photo-sensitization is a topic of much current scientific interest. Based on this, a study related to the TNTs' sensitization process with methylene blue was accomplished. Considering the structural and morphological differences, the TNT20012, TNT16024 and CoTNT20012 samples were selected for this study. Before the adsorption experiments, the BET surface area of the samples was measured (Table 2). The specific surface area values, $S_{BET}$, obtained for the nanotubular samples TNT16024 and CoTNT20012, were 205.55 and 194.55 $m^2 g^{-1}$, respectively. As expected, those values contrast with the considerable lower value of 21.08 $m^2 g^{-1}$ obtained for the heterogeneous nanorod-*like* TNT20012 sample. Concerning the adsorption affinity, the MB was adsorbed by all the tested samples, confirming that their surfaces are negatively charged when in contact with the cationic aqueous dye solution [28], and in agreement with the fact that the interactions involved in the adsorption phenomena are predominantly electrostatic [29]. The MB adsorption capacities of the CoTNT20012, TNT20012 and TNT16024 samples are given in Table 2. As can be seen from Table 2, considerable different dye amount were adsorbed by the different samples. The CoTNT20012 sample shows the highest MB adsorption values, with 213.29 mg of MB adsorbed per sample gram. Under identical experimental conditions, the MB adsorption drops to 137.54 mg g$^{-1}$ for the undoped TNT16024 sample and to 37.68 mg g$^{-1}$ for the undoped TNT20012 sample.

The different surface areas cannot explain by itself the difference in the adsorption results, since the $S_{BET}$ values for CoTNT20012 and TNT16024 samples are comparable, 194.55 and 205.55 m$^2$ g$^{-1}$, respectively, while the corresponding MB adsorption values are quite different, 213.29 and 137.54 mg g$^{-1}$, respectively. Therefore, the influence of the Co-doping process in the TNT's surface activity and/or in the crystalline structure interlayer space should also be considered to better understand these adsorption results.

The methylene blue molecule can be visualized as a rectangular shaped molecule, with dimensions of approximately 1.43 nm × 0.60 nm and thickness of 0.18 nm (values estimated using the *ChemSketch* 12.01 software). Since it is reasonable to assume that MB molecules adsorb at the TNTs' surface through electrostatic interactions as mentioned above, the MB molecules may lay perpendicularly oriented, with the positive charge towards the TNTs' surface, occupying an area of *ca.* 0.26 nm$^2$ per molecule. On the other hand, due to the delocalized $\pi$ system in the aromatic rings of the dye molecules, the flat-lying adsorption in the surface is also possible, resulting in this case an occupied area per MB molecule of *ca.* 0.86 nm$^2$. The area occupied by the total adsorbed MB, $S_{MB}$, considering these two possible coverage extremes (MB molecules adsorbing perpendicular or parallel to the surface), were estimated and are presented in Table 2. As can be seen from the data, the specific surface areas of TNT20012, TNT16024 and CoTNT20012 samples are lower than the ones calculated for the highest MB covering (flat-lying orientation), suggesting that at least part of the dye could be intercalated in the interlayer space of the titanate lamellar structure [30]. On the other hand, the occurrence of MB intercalation between the TiO$_6$ layers is possible for the TNT and CoTNT since they possess sodium ions in the interlayer space, available to be exchanged. This seems to be a reasonable assumption since the MB molecule dimensions and the TNTs' interlayer distance reported in literature (~0.9 nm) [10,30] are compatible with the MB intercalation occurring with the molecule in a parallel position to the TiO$_6$ layers. However, note that, from the analysis of the data presented in Table 2, if MB intercalation in the TNT16024 sample occurs it should be marginal. For the CoTNT20012 the dye intercalation process must be taken into account.

### 3.4. Electrochemical characterization

Methylene blue is an electroactive molecule [31], and therefore, the dye-sensitized materials were electrochemical characterized in order to study their redox behaviour, which would validate the dye molecule intercalation hypothesis proposed in section 3.3. Two samples were selected for the electrochemical study: TNT20012 and CoTNT20012. They were previously saturated with MB and supported in the GCE electrode surface, giving rise to the systems labelled MB-TNT20012/GCE and MB-CoTNT20012/GCE, respectively. For comparison purposes, the MB redox potential process was studied using a GCE electrode immersed in a MB aqueous solution (MB/GCE system) while the redox process of the TNT20012/GCE and CoTNT20012/GCE systems (*non*-sensitized systems) were studied using a blank electrolytic solution. Figure 6 shows the respective cyclic voltammograms and Table 3 summarizes the obtained electrochemical data. The voltammogram of the MB/GCE system shows a reduction peak at a potential $E_{red} = -324$ mV, which is attributed to the *two*-electron reduction of the MB cation (MB$^+$) to the neutral *leuco*-MB form [32]. The oxidation of this *leuco*-MB form occurs at a potential of $E_{ox} = -278$ mV. On the other hand, when MB is adsorbed on the TNT20012 surface (MB-TNT20012/GCE system), the dye reduction occurs at $E_{red} = -249$ mV, a lesser negative potential than the value in aqueous solution (see Table 3). The MB-CoTNT20012/GCE system has a reduction potential of $E_{red} = -301$ mV. This value is higher than the one obtained for the MB/GCE system (−324 mV). On the other hand, the MB-CoTNT20012/GCE cathodic peak potential is more negative than the one obtained for the undoped MB-TNT20012/GCE system (–249 mV). Additionally, it can be seen that the redox process reversibility of the MB-CoTNT20012/GCE system is lower (higher $\Delta E$) comparatively to the MB-TNT20012/GCE and MB-GCE systems, which indicates that the presence of Co in the TNT structure affects the dye electrochemical response. Assuming that a higher overpotential will be required for the redox process of more stabilized MB molecules (stronger immobilization), and taking into account that the dye molecules adsorbed in the TNT surface will be more easily oxidized and reduced than the intercalated ones, the lower $E_{red}$ value and the lower electrochemical reversibility observed for the MB-CoTNT20012/GCE system (when compared with MB-TNT20012/GCE) supports the hypothesis of a higher MB intercalation degree in the Co-doped sample.

## 3.5 MB-CoTNT optical behaviour and stability

In order to study the stability of the dye sensitized TNT materials under radiation, the MB-sensitized samples were submitted to intense UV irradiation. Figure 7a shows images of the MB-TNT16024, MB-TNT20012 and MB-CoTNT20012 powder samples before and after being submitted to strong UV continuous irradiation during 2 hours. As can be seen, the samples' colour intensity is in accordance with the different amounts of MB adsorbed by each material (see Table 2): lowest for the TNT20012 sample and highest for the CoTNT20012 one. On UV irradiation, the colour of the powders changes. The change in colour is clearly perceptible for the undoped TNT20012 and TNT16024 samples and almost indistinguishable for the doped CoTNT20012 sample.

The reflectance spectra of the three dye-sensitized powders taken before and after UV irradiation are presented in Figure 7b. For comparison purposes, Figure 7b also displays the reflectance spectra of an aqueous MB solution. As can be seen, the MB absorption spectrum presents a strong absorbance band centred at 663 nm, which is characteristic of the monomer form of the dye ($MB^+$). The shoulder at 617 nm band depends on the dye concentration and is attributed to dimeric or slightly larger aggregates of the MB molecules. On the other hand, the absorption spectrum of MB-CoTNT20012 sample starts at 900 nm (near IR) and is extended in the visible and UV range, with a broad absorption peak centred at ~600 nm. Note that this spectrum is rather different from the one of the corresponding sample before being dye sensitized (see Figure 3) and cannot be understood as a simple convolution of both MB and CoTNT20012 sample absorption spectra. Therefore, the absorption behaviour of sample MB-CoTNT20012 can only be due to the strong interactions between the adsorbed MB molecules and the titanate nanostructure. The broadening of the peak at ~665 nm indicates the co-existence of MB molecules and different type of dye agglomerates (monomeric, dimeric and trimeric) [33]. After two hours of UV irradiation, the profile of the absorption spectrum was preserved, although a slight decrease of the total absorbed energy has occurred.

A similar behaviour was observed for the MB-TNT16024 sample. In this case the absorption edge starts at a lower wavelength (~850 nm) and the total absorbed energy is lower in comparison to the energy absorbed by the MB-CoTNT16024 sample. This suggests the existence of a lower amount of active MB molecules in the MB-TNT16024 sample, in accordance with the qualitative information given by the corresponding images of Figure 7a. The radiation absorption of the nanorod-*like* MB-TNT20012 sample is very low and their absorbance spectra before and after irradiation are almost feature-*less* (not shown).

4. Conclusions

Cobalt-doped titanate nanotubes with a high length/diameter aspect ratio were successfully prepared trough a hydrothermal synthesis method using an amorphous Co-doped precursor. It was shown that Co-doping stabilizes the morphology of titanate nanotubes in different synthesis temperatures, 160 ºC and 200 ºC, and significantly reduces the autoclave dwell time. Concerning the optical behaviour of the synthesised materials, it was shown that the band gap energies of the Co-doped TNTs are red shifted in relation to the energy values deduced from the undoped TNTs, with the doping process shifting the absorption edge of those samples to the visible region. The correlation between the samples' band gaps and their Urbach energies seems to support that the aforementioned red shifts are due to band-to-tail and tail-to-tail transitions, driven by the Co-doping and consequent generation of oxygen vacancies. The Co-doped samples display the highest dye immobilization ability, possibly due to the intercalation of the methylene blue molecules between the $TiO_6$ layers of the TNTs' structure, in addition to the surface adsorption. Finally, it was shown that after sensitization with methylene blue, the absorbance of both undoped and Co-doped TNT samples were extended from the near IR to the UV, with a strong absorption in the visible region. In particular, the dye sensitized CoTNT20012 sample exhibited a strong stability under UV irradiation and showed the highest absorbance in a wider region of wavelengths. This last result suggests that hybrid MB-CoTNT materials could have potential applications as photoactive materials in a broad range of the electromagnetic spectrum.

**Acknowledgments**

This work was supported by Fundação para a Ciência e Tecnologia under contract PTDC/CTM NAN/113021/2009. O.C. Monteiro acknowledges PEst-OE/QUI/UI0612/2011 and Programme Ciência 2007.

**TABLE CAPTIONS**

**Table 1**. Average diameter, optical band gap energy and Urbach energy of the different TNT samples.

**Table 2**. BET surface areas and MB adsorption amounts for different TNT samples.

**Table 3**. Electrochemical data determined for the MB-GCE, MB-TNT20012/GCE and MB-CoTNT20012/GCE systems.

**FIGURE CAPTIONS**

**Figure 1**. a) XRD pattern of the undoped TNT16024 sample. b) TEM micrograph of the undoped sample TNT16024; the inset displays the distribution histogram of the nanotubes' diameters. c) XRD pattern of the doped CoTNT16024 sample. d) TEM micrograph of doped CoTNT16024 sample; the inset displays the distribution histogram of the nanotubes' diameters.

**Figure 2**. a) XRD pattern of the undoped TNT20012 sample. b) TEM micrograph of the undoped TNT20012 sample; the inset displays the distribution histogram of the elongated nanostructures' diameters. c) XRD pattern of the doped CoTNT20012 sample. b) TEM micrograph of the doped CoTNT20012 sample; the inset displays the distribution histogram of the nanotubes' diameters.

**Figure 3**. Diffuse reflectance spectra of the different prepared TNT samples.

**Figure 4**. Tauc plots for the different TNT samples synthesized. The optical band gap energies were estimated by extrapolating the linear portion of the curve to zero absorption.

**Figure 5**. a) graphic representation of $\ln(F_{KM})$ *vs.* photon energy, from which the Urbach energy of the samples was evaluated. b) optical band gap energy *vs.* Urbach energy for the undoped and Co-doped samples. Each data point is labelled with the corresponding sample reference.

**Figure 6**. Cyclic voltammograms for the MB/GCE, MB-TNT20012/GCE and MB-CoTNT20012/GCE systems. Insert: cyclic voltammograms for the TNT20012/GCE and CoTNT20012/GCE systems in blank electrolytic solution.

**Figure 7**. a) Powder images and b) absorption spectra of the MB-TNT and MB-CoTNT samples before and after 2 hours of UV radiation. The reflectance spectra of an aqueous MB solution it is also displayed for comparison purposes.

**Table 1**

| Sample | <*d*> (nm) | $E_g$ (eV) | $E_U$ (eV) |
|---|---|---|---|
| TNT16024 | 8 ± 2 | 2.96 ± 0.03 | 0.578 ± 0.005 |
| TNT20012 | 110 ± 81 | 3.44 ± 0.03 | 0.502 ± 0.116 |
| CoTNT16024 | 7.1 ± 0.8 | 2.37 ± 0.05 | 1.555 ± 0.352 |
| CoTNT20012 | 7 ± 2 | 2.36 ± 0.04 | 1.595 ± 0.642 |

**Table 2**

| Sample | $S_{BET}$ (m$^2$ g$^{-1}$) | MB adsorption (mg g$^{-1}$) | $S_{MB}$ (m$^2$) | |
|---|---|---|---|---|
| | | | *Lower*[a] | *Higher*[b] |
| TNT16024 | 205.55 | 137.54 | 67.12 | 222.89 |
| TNT20012 | 21.08 | 37.68 | 18.38 | 60.94 |
| CoTNT20012 | 194.55 | 213.29 | 104.08 | 344.89 |

[a] Lower – lower value of occupied area, considering MB molecules lying with the positive charge towards the TNTs' surface.
[b] Higher – higher value of occupied area, considering MB flat-lying molecules on the TNTs' surface.

**Table 3**

| Electrode | $E_{ox}$ (mV) | $E_{red}$ (mV) | $\Delta E$ (mV) |
|---|---|---|---|
| MB-GCE | −278 | −324 | 46 |
| MB-TNT20012/GCE | −198 | −249 | 51 |
| MB-CoTNT20012/GCE | −198 | −301 | 103 |

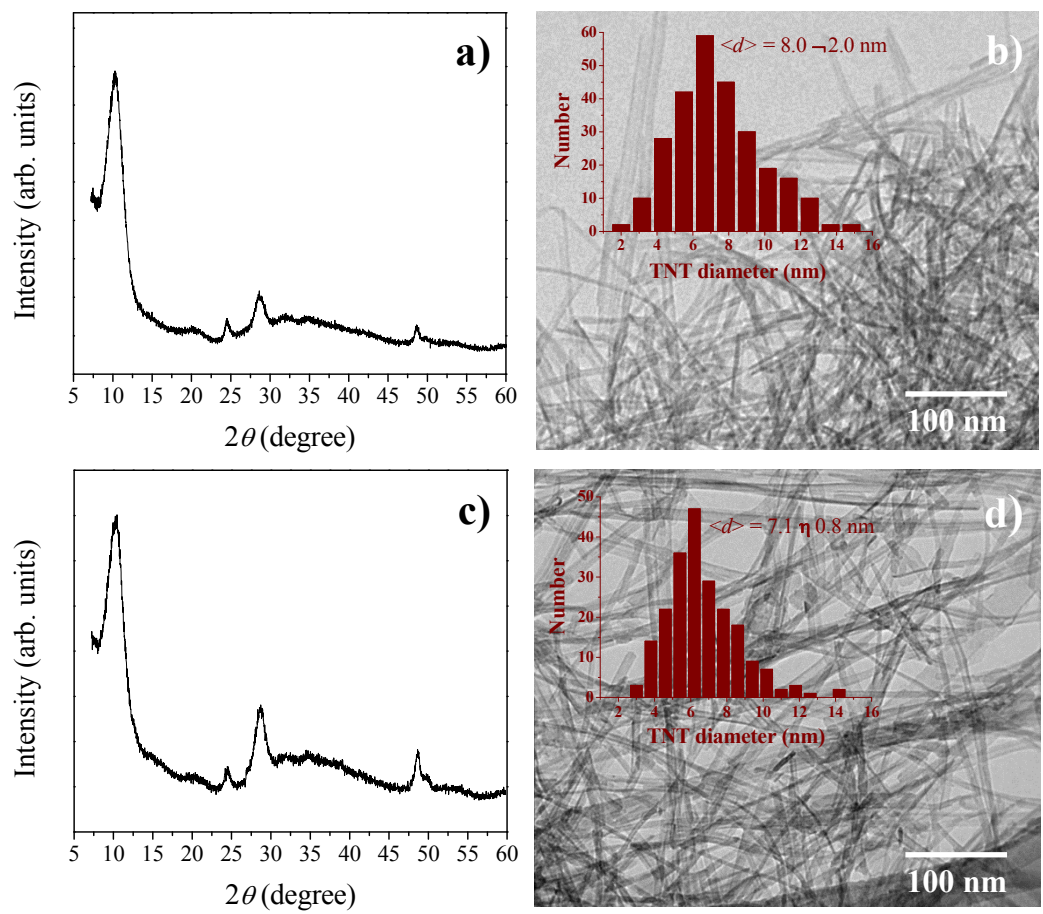

**Figure 1**

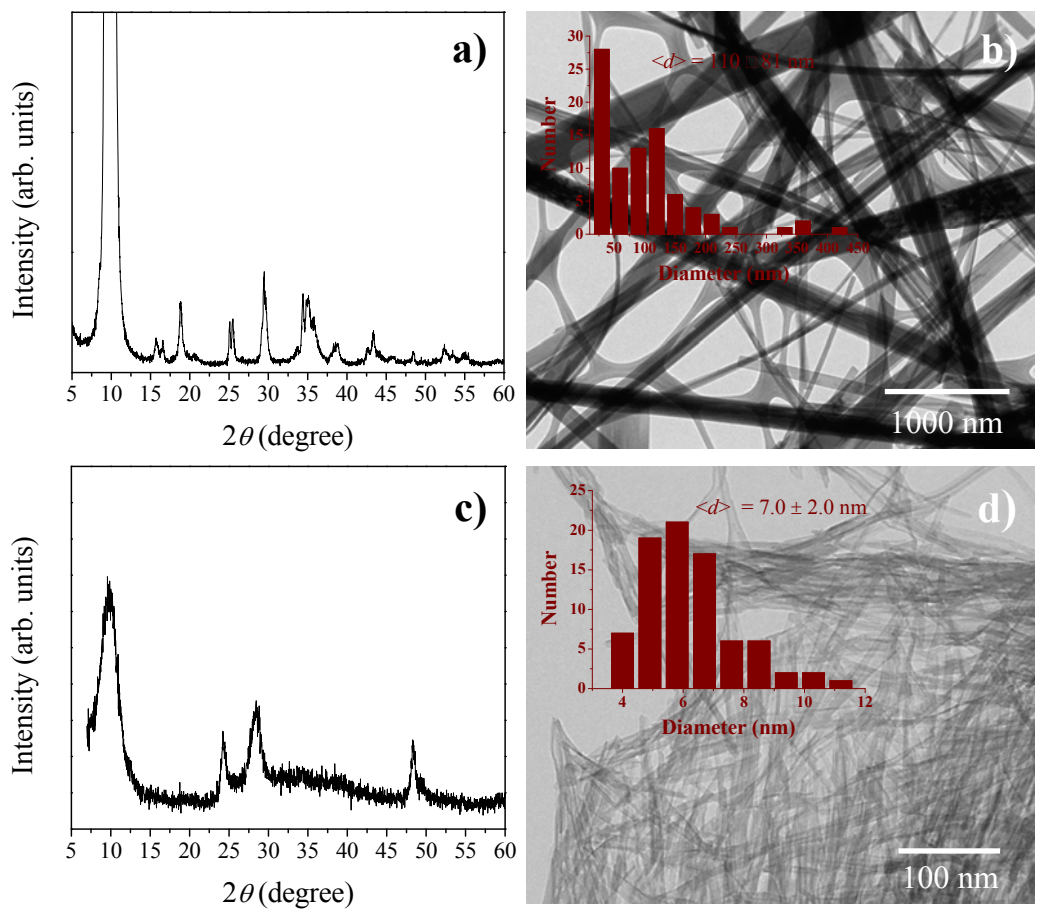

**Figure 2**

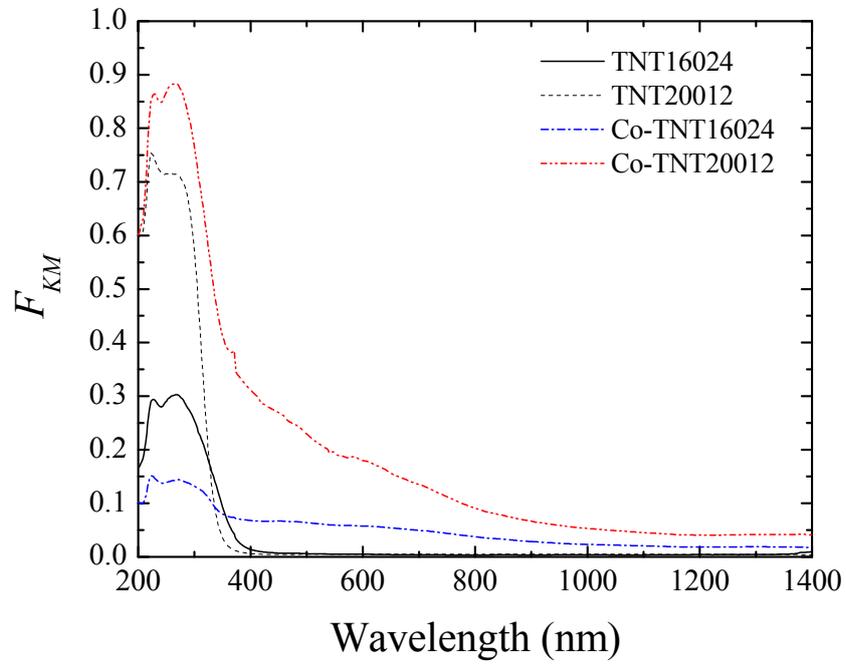

**Figure 3**

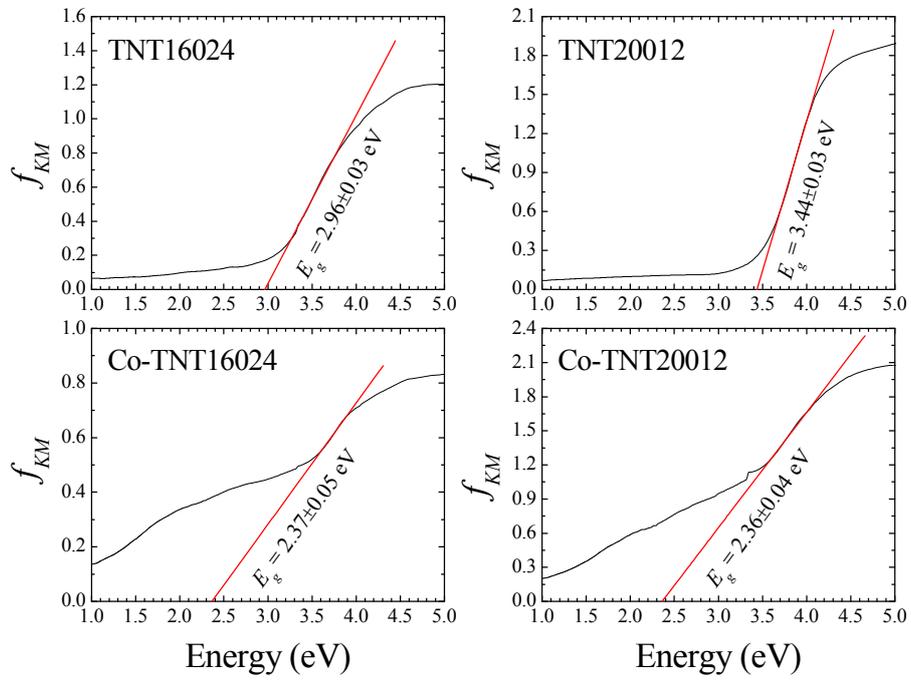

**Figure 4**

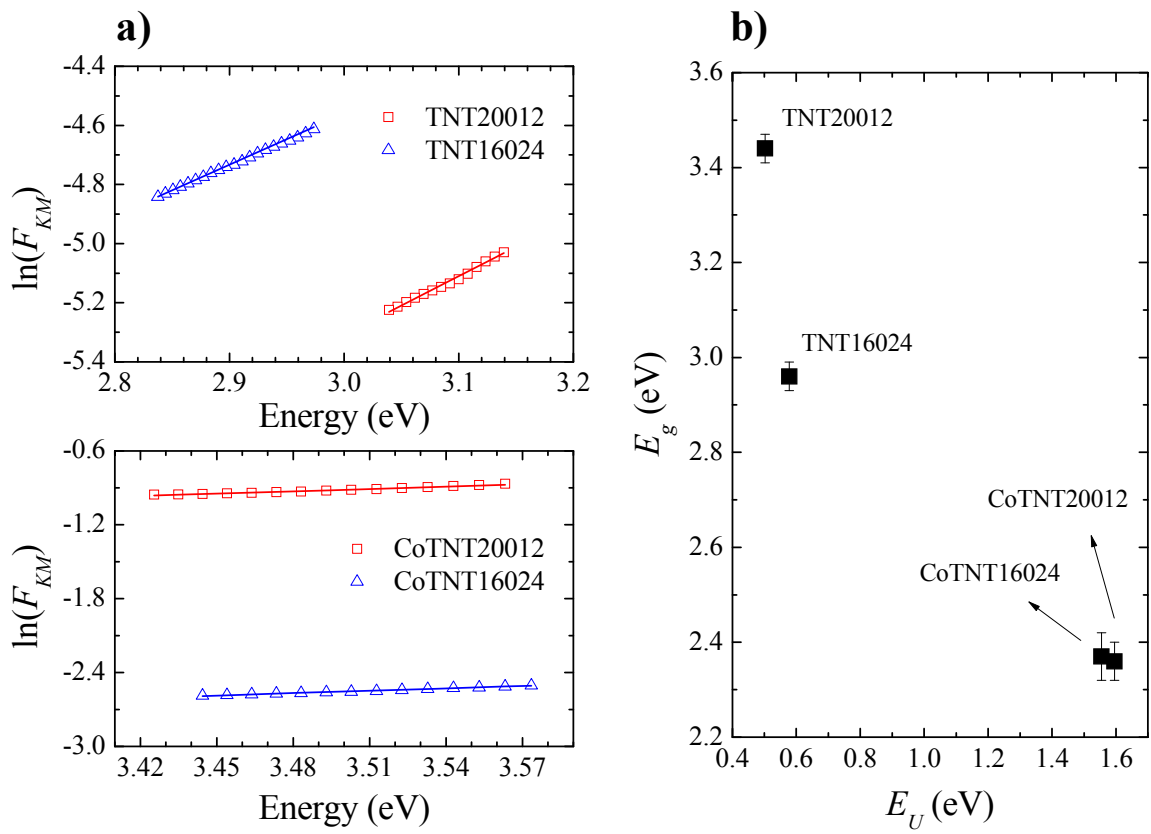

**Figure 5**

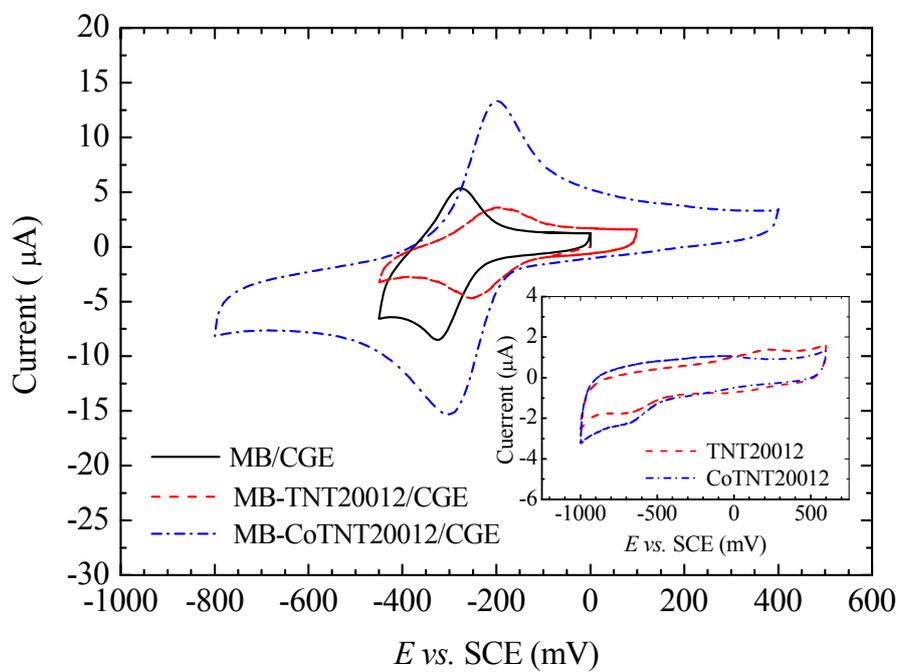

**Figure 6**

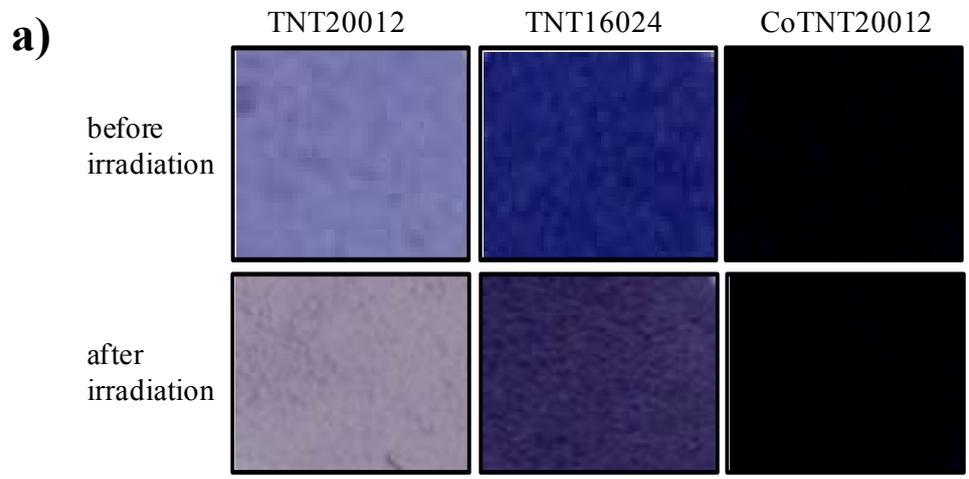

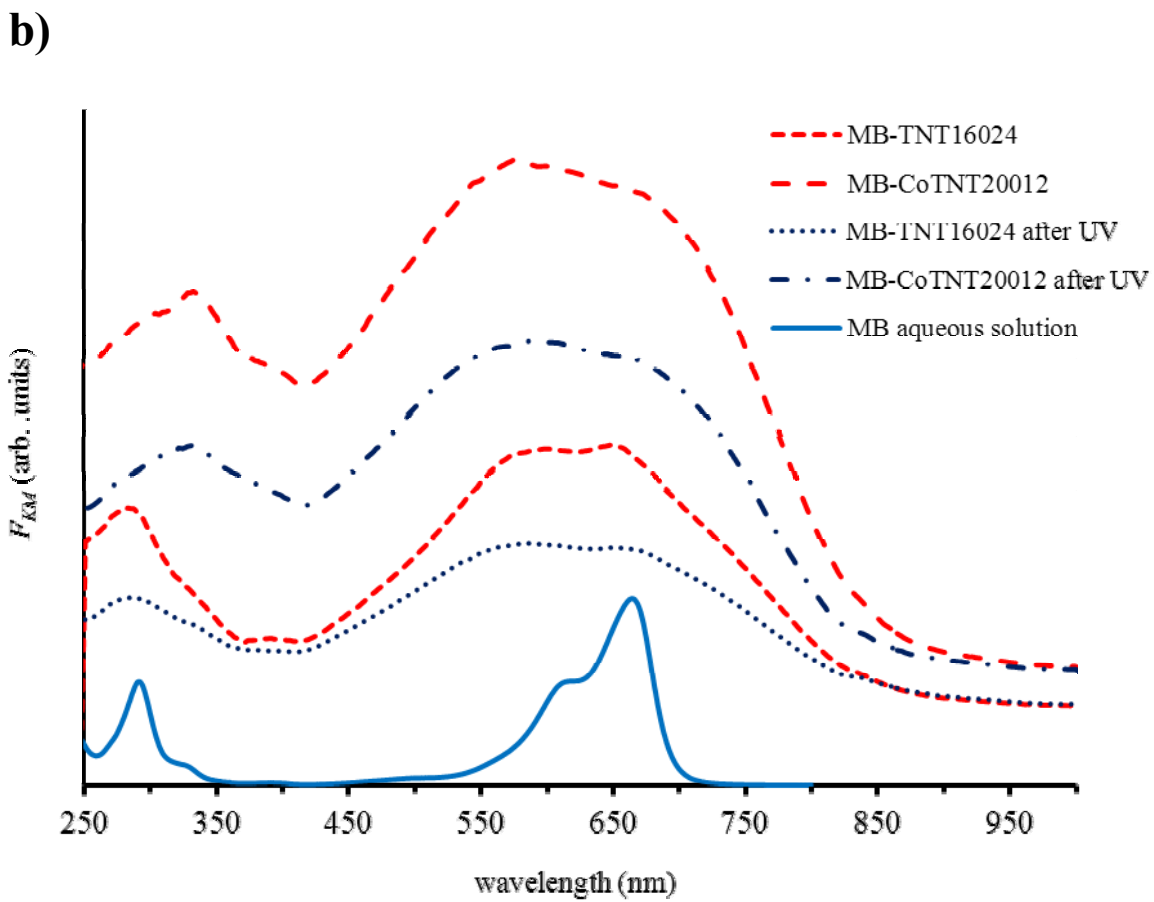

**Figure 7**

# Electronic supplementary data

# Synthesis and properties of Co-doped titanate nanotubes and their optical sensitization with methylene blue

V.C. Ferreira[1] M.R. Nunes[2], A.J. Silvestre[3] and O.C. Monteiro[2*]

[1] University of Lisbon, Faculty of Sciences, Department of Chemistry and Biochemistry and CQB, Campo Grande, 1749 – 016 Lisbon, Portugal

[2] University of Lisbon, Faculty of Sciences, Department of Chemistry and Biochemistry and CCMM, Campo Grande, 1749 – 016 Lisbon, Portugal

[3] Instituto Superior de Engenharia de Lisboa, Department of Physics and ICEMS, R. Conselheiro Emídio Navarro 1, 1959-007 Lisboa, Portugal

**Energy dispersive X-ray spectroscopy (EDS) data**

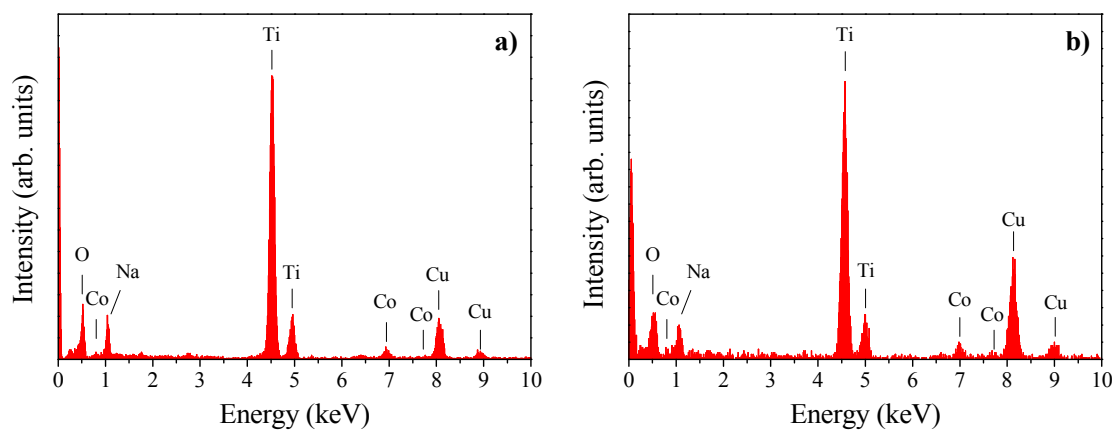

**Figure S1.** EDS spectra of a) doped CoTNT16024 and b) CoTNT20024 samples. Quantitative analyses allowed to estimate Co:Ti ratios of 4.4% and 4.3% for samples CoTNT16024 and CoTNT20024, respectively. The Cu peaks assigned are due to the copper grid used.

[2] Corresponding author: Phone:+351 217500865; Fax:+351 217500088; E-mail: ocmonteiro@fc.ul.pt